# 自动驾驶安全等级及定位精度设计

唐恬恬
许浩
吴诚成
吕思婕
项艳

## 摘要

自动驾驶是当前科技前沿的研究热点，科技公司和传统车企正在从两个不同方向开发设计自动驾驶技术。本文将从自动驾驶分类标准以及 ISO 安全等级出发，结合中国交通事故数、死亡数据等，并参考美国对自动驾驶虚拟驱动系统的风险分配方法进行中国虚拟驱动系统的风险分配。此外，结合车辆"定位盒子"模型进行中国定位精度警报限制理论计算并进行相关车型定位精度需求设计。

**关键词：** 自动驾驶,安全等级,定位精度

## Abstract

Autonomous driving is a hot research topic in the frontier of science and technology. Technology companies and traditional car companies are developing and designing autonomous driving technology from two different directions. Based on the automatic driving classification standard and ISO safety level, combined with the number of traffic accidents and death data in China, and referring to the risk allocation method of the automated driving virtual drive system in the United States, the risk allocation of China's virtual drive system will be carried out. In addition, combined with the vehicle "positioning box" model, the theoretical calculation of the alarm limit of positioning accuracy in China will be carried out and the positioning accuracy requirements of related vehicles will be designed.

# 1. 绪论

1.1 国内外研究现状

## 1.1.1 自动驾驶相关行业发展现状

从自动驾驶概念被提出以来,人们就对它未来的发展期待。而随着自动驾驶多年来的发展与制定的规范,由国际汽车工程师协会制定的自动驾驶分级标准[19],即如下表所示。

表 1 SAE 等级划分表[1]

| 自动驾驶分级 | 等级名称 | 控制主体 | 设计内容和条件 | 特殊附加条件 |
| --- | --- | --- | --- | --- |
| L0 | 无自动化 | 驾驶员 | 驾驶员全权操作汽车 | 无 |
| L1 | 部分驾驶辅助 | 驾驶员和辅助系统 | 在 L0 基础上添加操作提供系统 | 辅助驾驶 |
| L2 | 组合驾驶辅助 | 驾驶员和车辆系统 | 添加环境对方向盘和减速中多项操作 | 辅助驾驶 |
| L3 | 有条件自动驾驶 | 车辆系统 | 在合适的情况下由系统完成所有的驾驶操作 | 系统控制转向和加速度,驾驶员变为乘客 |
| L4 | 高度自动驾驶 | 自动驾驶系统 | 由系统完成所有驾驶操作 | 除少数特殊环境情况 |
| L5 | 完全自动驾驶 | 自动驾驶系统 | 任何路况下均为系统驾驶 | 预测需要等到强 AI 时代 |

此外,我国即将实行的分级与之类似。该分级自 2017 年启动,由长安汽车、中国汽车技术研究中心、吉利、广汽等九家企业根据我国的实情分为 0-5 六个等级,历时三年完成,并由国家工信部发布并报批公示,名为《汽车驾驶自动化分级》[2]。

目前自动驾驶商业化已经发展出泾渭分明的两个方向,一是 IT 派,二是车企派。首先介绍 IT 派,IT 企业以 RoboTaxi 为载体研发 L4/L5 级别自动驾驶,代表公司有 Google,Waymo,小马智行,Momenta,百度 Apollo 等多家科技公司,这些高科技公司由于有强大的科技研发能力,在"软"的方面拥有得天独厚的优势,比如完善的软件开发、测试流程及人才储备并且没有量产的压力等。由于 IT 企业的最终目标是将自动驾驶相关技术分段打包出售,他们大多跳过基础的研究阶段,而寄托于通过全程的车联网,构建一种出行的便捷服务。本文选取几个具有代表性企业介绍他们当前已研发的技术。例如,谷歌的 Waymo,该公司研发的硬件产品,除传统的超声波雷达、自主研发的激光雷达、摄像头、自主研制的视觉传感器外,还使用了音频检测系统、GPS、IMU 等,并且搭载了车道保持系统;在软件技术部分,创建"ChauffeuNet"的深度循环神经网络(RNN),深度学习和人工智能为一体的技术,车轮角度编码器 Wheel Encoder 等。再如 Momenta 公司研发了基于深度学习的实时环境感知与高精地图,除辅助传感器雷达超声波雷达等传统设备外,致力于打造"飞轮式"L4。软件技术,有诸如量产数据、全流程的数据驱动算法、闭环自动化、自动驾驶决策算法等具特色的技术。Apollo 公司应用了工控机 IPC(包括 GPU)、GPS、IMU 惯性系统、CAN 总接线口卡、大容量硬盘、顶部旋转 Lidar、前、侧向摄像头、车头、车身 Lidar、超声波雷达等,研发了 Linus 内核、Novotel 的 GPS 和 IMU 组合定位系统,加入了 Deep-Drive 深度学习自动驾驶产业联盟等。再如小马智行也采用雷达、摄像头、GPS+IMU、轮速传感器等,通过研发动态

寻路算法，从而构建了 Ponybrain 自动驾驶基础架构平台，它能支持复杂的深度学习模型和软件快速迭代[3]。

  另一个方向是以量产车为基础，即以造车企业为代表的车企派别，这些企业在现有车辆基础上加装自动驾驶辅助系统，服务于驾驶者，并拥有完整的造车产业链，故而主要研发 L2 及 L3 级自动驾驶。由于他们的终极目标是销售车辆，自动驾驶便成为车辆销售中最大的附加值。虽然目前许多车企宣传他们的产品达到 L2.5 甚至是 L3 级别，实际上大部分量产的汽车都处于 L2 级别。本文同样选取几个企业的代表产品简单介绍他们应用的技术，例如，理想 ONE，车辆搭载摄像头数量是 5 个，前置辅助驾驶感知摄像头 1 个，毫米波雷达 1 个，超声波雷达 12 个，自动驾驶芯片是 EyeQ4，芯片算力为 2.5Tops，自动驾驶辅助系统是 ADAS。非常热门的特斯拉 model 3，搭载的摄像头数量是 8 个，前置辅助驾驶感知摄像头 3 个，毫米波雷达 1 个，超声波雷达 12 个，自动驾驶芯片为 HW3.0，芯片算力为 144Tops，自动驾驶辅助系统是 Navigate on Autopilot（NOA）。以及蔚来 ES6，摄像头数量是 7 个，前置辅助驾驶感知摄像头 3 个，毫米波雷达 5 个，超声波雷达 12 个，自动驾驶芯片是 EyeQ4，芯片算力为 2.5Tops，自动驾驶辅助系统是 Navigate on pilot（NOP）。还有小鹏 P7，摄像头数量是 13 个，前置辅助驾驶感知摄像头 1 个，毫米波雷达数量是 5 个，超声波雷达是 12 个，自动驾驶芯片 Xavier，芯片算力为 30Tops，自动驾驶辅助系统是 Navigate Guided pilot（NGP）[4]。

  当前自动驾驶行业总体是蓬勃发展的，也正在朝全面商用的目标快速推进。但现阶段的自动驾驶技术还不够成熟，距商用还有一段距离，仍然存在许多难关需要去攻克，比如激光雷达造价昂贵，高精地图实时更新较困难且耗资巨大等。此外，目前商用定位技术水平只能使自动驾驶汽车最高时速达到 40-60km/h，而且在车辆转弯时还会出现出现偏离车道的问题。

**1.1.2 应用于自动驾驶的 GNSS 定位技术研究现状**

  从上述中可以看出，当前自动驾驶技术众多，相关高精度定位技术在很大程度上影响着自动驾驶等行业的发展，也逐渐成为人们的研究热点。GNSS，即 Global Navigation Satellite System，全球卫星导航系统，在自动驾驶的高精度定位环节起到了最为关键的作用。目前现有四个全球卫星导航系统，分别是美国的 GPS、俄罗斯的 GLONASS、欧盟的 GALILEO 和中国的 BDS。[20]

  如果只借助 GNSS 系统的伪距观测值，最多只能实现米级定位，而为了达到更高的精度，则需要在 GNSS 定位的基础上结合其他方法以形成更优化的定位技术。在这一方面，当前广泛运用的定位技术有实时动态定位技术( real time kinematic，RTK) 精确单点定位技术 ( precise point position， PPP)。

  实时差分定位技术（Real Time Kinematic, RTK）指利用基准站和流动站之间系统误差的相关性，实时处理两个测站载波相位观测质量，将基准站采集的载波相位数据发送给用户流动站[8]，通过差分来消除和减弱相应的误差源，进而求差解算流动站相对基准站的坐标，实现高精度定位。常规 RTK 的有效作业距离在 20km 以内，一旦流动站和基准站的距离超过 50 km，常规的 RTK 的定位精度严重下降到分米级别。在此基础上发展到网络 RTK，可以将网络扩展到 100 km，网络 RTK 作业流程是首先流动站向服务端发送自身的大致位置，然后服务端根据用户流动站所在的位置生成一组虚拟观测值数据，通过数据传输协议将基于观测值域数据（OSR）传输给用户流动站，以便于用户流动站进行差分定位。核心在于参考站坐标精确已知，从而可以确定基准站间的电离层、对流层等空间相关误差项[8]。目前，RTK 技术的定位精度已经可以达到厘米级[7]。虽然定位精度较高，但 NRTK 具有以下缺点：1.依赖于密集的基准站资源；2.多个 CORS 区域存在盲区时无法连续性服务；3.需要频繁双向交互

通信；4.由于传输是基于观测值域，对通信宽度要求高。

精密单点定位技术（Precise Point Positioning, PPP）指利用高精度的卫星轨道和钟差，综合考虑各项误差模型的精确改正，通过非差方式解算单台 GNSS 接收机采集的相位和伪距观测值，获得统一框架下的高精度坐标。该技术摆脱对参考站的依赖，只需单台接收机，单向传播星历信息，即可实现全球性的高精度定位，因此具有成本低，作业灵活，且仅需单向数据传输等优点，对实现有无参考网络下的无缝切换，具有重要的应用价值。然而，如何高效快速的初始化能力是精密单点定位面对的一个重大挑战。

RTK 技术和 PPP 技术都日趋成熟并被广泛地应用在各个行业当中。但在应用的同时，网络 RTK 技术受限于数据通信负担较大且对覆盖范围的要求较高，而 PPP 技术的初始化精密定位时间很长。这些缺点都极大地限制了 RTK 和 PPP 技术进一步推动自动驾驶中导航的发展。PPP-RTK 技术的提出使得这一困境出现了转机。PPP-RTK 技术指以 PPP 为定位模式，加入区域相关的电离层和对流层增强信息，通过参数域（SSR）的增强方式，实现快速高精度定位的方法。PPP-RTK 技术具有以下优点：由于 PPP-RTK 采用的 SSR 增强信息模式，可以根据参数的物理特性来采用不同的播发频率，因此相对于 RTK，PPP-RTK 可以降低数据播发量，另外 SSR 模式将不同增强信息进行分开播发，也可针对性的提供完好性信息和快速发现增强信息中存在的问题；由于采用的是 PPP 定位模式，在有无参考站的情况下，PPP-RTK 可以实现无缝定位切换。

表 2 RTK、PPP 和 PPP-RTK 的对比

| | *RTK* | *PPP* | *PPP-RTK* |
|---|---|---|---|
| *精度* | 厘米-毫米 | 分米-厘米 | 厘米 |
| *收敛时间* | 瞬时 | 5-30 分钟 | 5 秒-1 分钟 |
| *覆盖范围* | 20-100 公里 | 大于 1000 公里 | 大于 100 公里 |
| *最大服务范围* | 局部 | 全球 | 全球 |
| *增强信息格式* | OSR | SSR | SSR |
| *通信方式* | 双向 | 单向 | 单向 |
| *所需带宽* | 中等 | 低 | 中等 |
| *增强信息* | 综合的观测值信息 | 轨道、钟差、偏差 | 轨道、钟差、偏差、对流层、电离层 |

从表格 2 可以看出，虽然 RTK 技术目前是比较成熟的高精度定位模式，可快速实现厘米到毫米级别的定位，但是由于双向通信和高并发问题，存在用户数据的隐私问题，并且增强信息是综合的观测值信息，很难针对不同参数给出特定的完好性信息，而完好性信息对自动驾驶的安全性至关重要。PPP-RTK 则结合了 RTK 和 PPP 的优点，很好的避免了这些挑战，作者也相信 PPP-RTK 是未来自动驾驶解决 GNSS 高精度定位更好的技术方案。

## 1.2 主要研究内容及章节安排

本文研究的是自动驾驶汽车的能够达到的安全级别以及所需的定位精度，主要分为三个部分，第一部分绪论包含当前自动驾驶行业现状，并介绍了 GNSS 中用到的 RTK 与 PPP 技术目前能够达到的定位精度性能，目的是为后续自动驾驶安全等级 TLS 验证和设计定位精度提供参考标准，以便检验自动驾驶安全是否达标以及当前技术是否足以支撑自动驾驶全面商用。第二部分是基于 ISO26262 安全等级和 TLS 对自动驾驶虚拟驱动系统进行风险分配。

第三部分，是由风险分配的数据进一步研究，设计确定自动驾驶汽车的定位所需的精度。

## 2. 基于中国道路交通数据的风险分配

在自动驾驶的发展过程中，安全性一直是人们关注的重点与研究的难点，由于安全事故中人为因素的占比很高，自动驾驶的推广被认为可以在一定程度上提高道路交通安全等级。以下部分将基于参考文献，从对自动驾驶的目标安全等级推导到虚拟驱动系统的风险分配，并结合中国数据提出适合本国道路交通情况的分析。

### 2.1 ISO 安全等级

随着自动驾驶的发展，电子电气系统在汽车领域发挥着越来越重要的作用，同时它对汽车安全的影响也与日俱增。如何系统化电子电气对安全的影响和预防系统性失效成为了当今驾驶行业不可回避的问题。

而对此问题的研究需要从基础的汽车等级评估出发，即 ASIL 安全等级。ASIL 安全等级，全称汽车安全完整性等级（Automotive Safety Integration Level），通常被用来描述系统能够实现指定安全目标的概率高低。ASIL 安全等级分为 4 个等级，即 A，B，C，D，其中 A 是最低等级，D 是最高等级[10]。

划分 ASIL 安全等级需要首先通过风险评估和危险分析，来初步决定等级的基本范围，然后逐级分解安全等级和要求，直至无法进一步分解，并验证分配的合理性，从而确定最终的定级。ASIL 安全等级的三个基本要素分别为严重度（Severity）、暴露率（Exposure）和可控性（Controllability）。

严重度，即对人员、财产将遭受损害的程度，比如说如果只是门锁故障，它的严重度就会比刹车故障低。我们用 SX 来表示它，并且将它分为四个等级，分别是：S0 无伤害；S1 轻伤；S2 重伤；S3 致命伤害。暴露率，即对人员或财产造成影响或干扰的概率，比如车顶出现轻微异响的暴露率会比司机驾驶位故障的暴露率低。我们用 EX 来表示它，并且将它分为五个等级，即从 E0 到 E4，E0 是几乎不可能暴露于危险中，E4 是可能性极高。可控性，即驾驶员等在多大程度上可以采取主动措施避免损害的发生，在车道上行驶时，轮胎缓慢漏气的可控性会比刹车故障的可控性高。用 CX 来表示它，并且将它分为四个等级，即从 C0 到 C3，最低 C0 可控，最高 C3 几乎不可控。

具体的评估结果范例表可以参考表 3。其中 IEC-61508 是电气/电子/可编程电子安全相关系统功能安全国际标准，DO-178/254 是用在航空或者发动机的机载系统和设备复杂电子硬件设计的质量保证导则，CENELEC 50126 128/129 则是有关轨道交通安全等级的标准。

表 3 安全等级划分范例表

| 每小时事故的可能性 | IEC-61508 | ISO 26262 | DO-178/254 | CENELEC 50126 128/129 |
|---|---|---|---|---|
| – | (SIL-0) | QM | DAL-E | (SIL-0) |
| $10^{-6}$-$10^{-5}$ | SIL-1 | ASIL-A | DAL-D | SIL-1 |
| $10^{-7}$-$10^{-6}$ | SIL-2 | ASIL-B/C | DAL-C | SIL-2 |
| $10^{-8}$-$10^{-7}$ | SIL-3 | ASIL-D | DAL-B | SIL-3 |
| $10^{-9}$-$10^{-8}$ | SIL-4 | – | DAL-A | SIL-4 |

## 2.2 风险分配理论

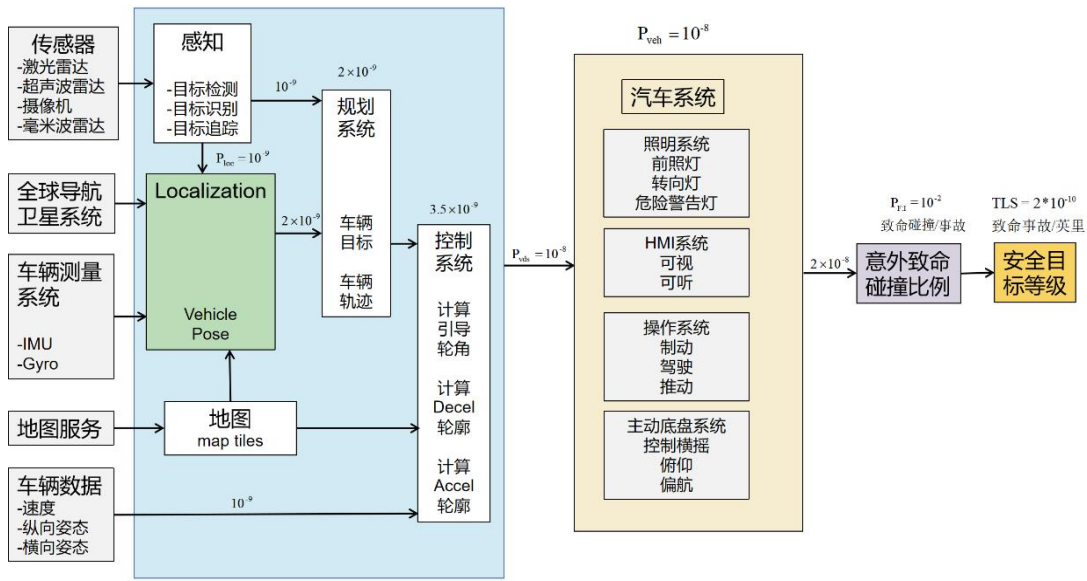

图 1 虚拟驱动系统风险分配流程[11]

首先,介绍安全目标等级 TLS (Target Level of Safety),并根据 TLS 的需求从右往左推导。由于航空的安全标准在最初建立时是把轨道交通的标准平移挪用过来的,现在也把航空安全标准平移挪用到汽车标准上。接下来采用美国的数据推导上图中设计流程。

从实际来看,航空行业的安全等级一直都处于最高水平,因此按照航空的级别来约束道路安全。当下航空安全级别为 $2.5×10^{-10}$,单位是每行驶公里的致命事故数,而当前道路安全的级别约为 $10^{-8}$,于是将道路安全目标设置为 TLS=$2×10^{-10}$,在当下的基础上提高两个数量级以对标航空的安全要求。

而按照美国道路交通事故数据来看,在 2016 年,交通事故死亡人数为 37,461 人,致命事故数为 34,439 起,可以得到 $\frac{死亡数}{致命事故数} = \frac{37,461}{34,439} = 1.09$,即每一场致命事故约有 1 人丧命。

在已知致命事故中死亡数的情况下,需要计算出致命事故占总交通事故的比例,即 $P_{F:I}$,而根据美国统计数据显示,在汽车交通事故中,比例约为 1:172,即在 172 场交通事故中会有 1 场致命事故,而在航空中,这个比例会有所上升,为 1:14,这是因为航空事故的危险性会大于道路交通事故。

同时,由于道路事故统计中有些数据没有上报,汽车致命事故的比例需要一定的修正。基于历史数据,可以保守地将汽车数据取为 $P_{F:I}=10^{-2}$。在已知以上数据的基础上,完整性风险的可接受水平即可被计算出来。

汽车系统的 $P_{veh}$,虚拟驾驶系统的 $P_{vds}$,TLS 和 $P_{F:I}$ 的关系如下:

$$TLS = P_{F:I}(P_{veh} + P_{vds}) \tag{1}$$

由于 TLS 是预期想要达到的安全水平,可以将已有的理想安全级别数据带入,计算出 $P_{veh} + P_{vds} = 2×10^{-8}$ 事故/英里,然后将汽车系统和虚拟驾驶系统的风险平均分配,从而得到了 $P_{veh} = P_{vds} = 10^{-8}$ 事故/英里。

要实现虚拟驱动程序系统 $P_{vds}$ 所需的完整性风险,仔细研究其子系统。在上述推导中,本文更多的专注于定位,而定位是由虚拟驱动程序系统的其他元素决定,这包括感知、定位、

规划和控制中的硬件和软件故障,这些因素综合在一起必须保证精准定位较高的成功率。图 1 显示了虚拟驱动程序系统的内部元素以及系统定位的重要性。定位的输出是对计划的输入,而计划的输出是对控制的输入,因此定位的失败会向下传播。图中同样显示了完整性风险对所有虚拟驱动子系统的一种可能分配,其中定位的目标是 $P_{loc} = 10^{-9}$ 事故/英里。

而根据真实数据计算,$\hat{P}_{veh} = \frac{5,800,000 \text{事故}}{3,005,829,000,000 \text{英里}} \times 2\% = 3.8 \times 10^{-8}$ 事故/英里,其中 2%是指由车辆系统导致的碰撞事故的比例。

尽管上面使用的每个值都存在不确定性,但误差是处于一个可接受的范围中,结果表明车辆系统处于安全标准建议的数量级[11]。

**2.3 中国道路安全等级分析**

与上述基于美国数据的推导过程类似,本文将再基于中国数据来推导具体的风险分配情况。调查发现,截至去年年底,我国民航首次实现亿客公里死亡人数十年滚动值为 0 和百万小时重大事故率十年滚动值为 0,所以不能直接以航空安全率为目标,我们引用文献中关于美国航空安全率的数据,也就是 $2.5 \times 10^{-10}$ 来推导中国具体相关的风险分配。由于目标安全等级并不会真正影响中国相关数据的计算,而是设置好"认为安全的度",所以引用美国的数据有一定的合理性。

同时,对于致命事故/道路事故数,中国在这方面并没有明确的可查询的数字,所以同样引用美国数据的 $10^{-2}$,在这里假设各个国家之间的差异较小,所以在这里使用美国数据作为参考计算不会带来很大误差。

关于中国的事故数和相应行驶公里数的比例,以 2018 年数据为标准,中国一年事故数量为 244,937 件,行驶公里数以每辆车一年公里数 1.5 万来计算,中国汽车在 2018 年的保有量为 $2.4 \times 10^8$ 辆[12]。

$$\hat{P}_{veh} = \frac{244,937 \text{ crashes}}{1.5 \times 10^4 \text{公里/辆} \times 2.4 \times 10^8 \text{辆} \times 0.621 \text{mile/公里}} \times 2\% = 2.2 \times 10^{-9} \text{ failures/mile}$$

通过这个数据,进行与美国相似的风险分配流程,计算得出了中国的 TLS=$2 \times 10^{-11}$ 致命事故/英里,这比预期的安全性还要高一个数量级,同样本文需要根据目标安全等级来反向分配整个驱动系统不同子模块的设计参数,由图一分配 $P_{vds}$ 子系统中的设计参数,控制模块为 $3.5 \times 10^{-10}$ 事故/英里,垂直方向数据模块为 $10^{-10}$ 事故/英里,PLANNING 模块为 $5.5 \times 10^{-10}$ 事故/英里,在 PLANNING 模块中,定位为 $10^{-10}$ 事故/英里,感知模块为 $10^{-10}$ 事故/英里,LOCALIZATION 共向 PLANNING 贡献 $2 \times 10^{-10}$ 事故/英里,每个模块权重根据他们重要程度决定。上述过程完成了对虚拟驱动系统的整个分配,也体现了中国的道路安全等级很高,能够很好地保证道路行驶安全。

**3. 定位精度警报限制范围计算**

在得到虚拟驱动系统的风险分配数据以后,根据实际情况中我国具体的车辆尺寸与道路尺寸的规定,即可计算出在保证安全等级的情况下对道路定位的警报限制值和允许范围内的定位误差,从而达到上述风险分配的应用目的。

**3.1 中国车辆尺寸简介**

我国有两套轿车尺寸体系,一套分为微型、小型、紧凑型、中型、中大型、大型,另一套分为 A00 级、A0 级、A 级、B 级、C 级、D 级,这两套分类标准基本上是一一对应的,

尺寸如下表 4 所示[13]。在探索自动驾驶车辆安全标准上，和美国一样，我国也形成了若干民间和半官方的智库，开展了一系列讨论和研究，不同科技公司对不同技术路线进行探索。德国是由整车制造商主导设计和测试，中国则是科技公司在自动驾驶的投资和研发上表现更活跃，而美国是同时对两条技术路线都进行了研究和快速地推进。

在开放路测上，美国做得最好，不但法律规定上明确了测试主体的责任，还对测试范围、周期和方式规划了完善的行政审批流程，使企业有法可依。而中国目前由于法律准备不足，导致了"盲测"（审批无据、责任主体义务不明确）现象的出现。

目前，国内仍没有迹象调整和出台车辆的安全标准，以适应自动驾驶技术的发展需求。我国《道路交通安全法》同样规定机动车驾驶员必须时刻把控方向盘。这些问题也有可能是由于我国道路情况标准比较复杂，不像国外足够简洁统一，需要未来更多的自动驾驶实验数据以及技术积累才能加以解决。

表 4 中国车辆尺寸简介[13]

| 车辆级别 | | 车长（总长）/毫米 | 车宽/毫米 | 轴距/毫米 |
|---|---|---|---|---|
| A00 | 微型 | 3400-3700 | 1600-1675 | 2300-2450 |
| A0 | 小型 | 3700-4000（两厢）4000-4400（三厢） | 1675-1750 | 2450-2600 |
| A | 紧凑型 | 4000-4300（两厢）4400-4700（三厢） | 1750-1825 | 2600-2750 |
| B | 中型 | 4700-5000 | 1825-1900 | 2750-2900 |
| C | 中大型 | 4900-5100 | 1860-1940 | 2825-1975 |
| D | 大型 | 5000-5300 | 1900-1975 | 2900-3050 |

### 3.2 水平定位警报限制范围计算

车辆的警戒边框（Bounding Box）在道路平面内的投影为一矩形边框，定义该矩形边框横向（lateral）的长度为 x，纵向（longitudinal）的长度为 y。为使定位精度在道路平面中达到车道级，则需保证车辆在道路平面内的矩形边框在各种情况下都处于一条车道的范围内；而按照道路平面内道路的形状，可以将道路分为直道和弯道两种。因此，车辆在道路平面内的矩形边框在直道和弯道两种情况下都应处于一条车道的范围内，如图 2 所示。

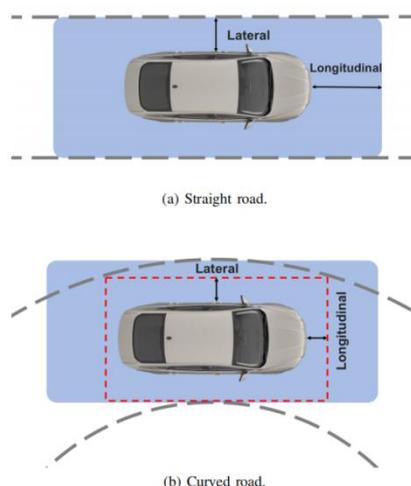

图 2 直道和弯道情况下的矩形边框[11]

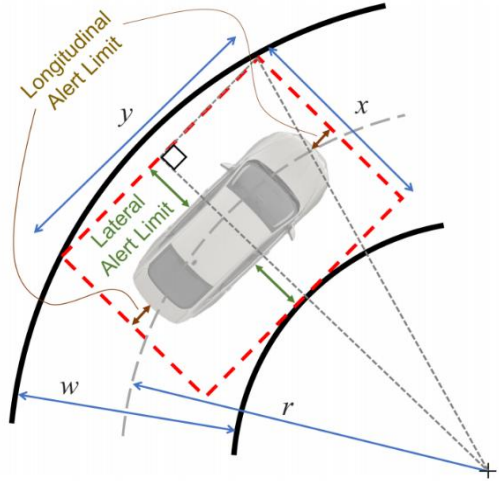

图 3 矩形边框恰好在车道内的极限情况[11]

其中与直道相关的几何参数为车道宽度，定义为 w。与弯道相关的几何参数为圆曲线最小半径，定义为 r，最大超高，定义为 i。公路弯道的超高是指当圆曲线半径小于不设超高的最小半径时，为抵消一部分横向力，将行车道绕旋转轴旋转，逐渐形成外侧高内侧低的单一横向坡度。在中国，弯道的圆曲线半径 r 和超高 i 满足公式：

$$r = \frac{v^2}{127(u+i)} \quad (2)$$

式中 v 为车速，u 为横向力系数[18]。

同时，弯道的最大超高与道路所处的情况有关，如表 5 所示[14]。

表 5 公路地域情况与弯道最大超高

| 公路所处地域情况 | 弯道最大超高 |
| --- | --- |
| 一般地区 | 8% |
| 积雪冰冻地区 | 6% |
| 以通行中小型客车为主的高速公路和一线公路 | 8%或 10% |
| 城镇区域 | 4%或 8% |

根据中国道路的设计要求[15]，整合以上所涉及到的所有道路几何参数，包括车道宽度 w、圆曲线最小半径 r 和最大超高，定义为 i，得到表 6。

首先讨论车辆在直道上的行驶情况，此时只需要求矩形边框的横向长度小于车道宽度，即 $x \leq w$。考虑到汽车速度在 40km/h 到 80km/h 之间的情况最多，取式中 w 为 3.75m 和 3.5m 中的较小值 3.5m，此时得到：$x \leq 3.5m$。

然后讨论车辆在弯道上的行驶情况，考虑矩形边框恰好在车道内的极限情况（如图 3），则由勾股定理，有 $\left(x + r' - \frac{w}{2}\right)^2 + \left(\frac{y}{2}\right)^2 = \left(r' + \frac{w}{2}\right)^2$ （3）

式中 $r'$ 表示车辆所在车道中轴线的圆曲线半径。

表 6 车道宽度 w 与弯道圆曲线最小半径 r 的数值均依赖于设计速度 v

| 设计速度 v(km/h) | 车道宽度 w(m) | 弯道圆曲线最小半径 r/m | | | |
|---|---|---|---|---|---|
| | | 最大超高 10% | 最大超高 8% | 最大超高 6% | 最大超高 4% |
| 120 | 3.75 | 570 | 650 | 710 | 810 |
| 100 | 3.75 | 360 | 400 | 440 | 500 |
| 80 | 3.75 | 220 | 250 | 270 | 300 |
| 60 | 3.5 | 115 | 125 | 135 | 150 |
| 40 | 3.5 | – | 60 | 60 | 65 |
| 30 | 3.25 | – | 30 | 35 | 40 |
| 20 | 3 | – | 15 | 15 | 20 |

考虑到 $r'$ 与 $r$ 接近，且 $r \gg w$，因此式中的 $r'$ 可由 $r$ 代替，即有

$$\left(x + r - \frac{w}{2}\right)^2 + \left(\frac{y}{2}\right)^2 = \left(r + \frac{w}{2}\right)^2, x > 0, y > 0 \tag{4}$$

简化后得变量 x,y 的显示函数式：$y = 2\sqrt{\left(r + \frac{w}{2}\right)^2 - \left(x + r - \frac{w}{2}\right)^2}, \quad x > 0$ (5)

根据表 6，在不同的最大超高下，同一设计速度对应不同的弯道圆曲线最小半径，因此参数 r 的取值情况比较复杂。为了简化计算，选取一般情况即最大超高为 8%（根据表 6）时对应的 r 进行考虑，相关数据如表 6 中蓝色区域所示。同时，考虑到一般情况下汽车行驶速度在 60km/h 到 80km/h 之间的可能性最大，因此选择设计速度分别为 60km/h 和 80km/h 的情况进行考虑，两种情况下(w,r)的值分别为(3.75m,250m)和(3.5m,125m)，函数图像如图 4 所示。

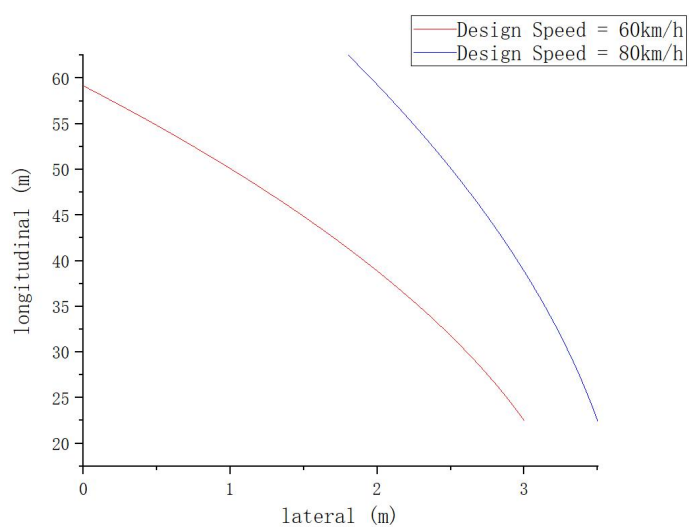

图 4 设计速度为 60km/h 和 80km/h 时的 x-y 函数曲线

根据图 4 可以看出，设计速度为 60km/h 时对 x,y 的要求更高。因此在设计速度为 60km/h 下计算横向和纵向的 alert limit。结合定义式：

$$\text{Lateral Alert Limit} = (x - w_v)/2$$
$$\text{Longitudinal Alert Limit} = (y - l_v)/2 \qquad (6)$$

并取中型轿车作为研究对象，在此情况下 $w_v = 1.8m, l_v = 4.7m$，得到图 5 所示函数曲线。同时根据公式(6)设计要求，Longitudinal Alert Limit 必须小于 1.5m，因此在图中作出 Longitudinal Alert Limit = 1.5m 的曲线，得到交点(0.82m,1.50m)，即在 Longitudinal Alert Limit 为 1.5m 的情况下，Lateral Alert Limit 为 0.82m。类似地，可以得到在其他设计速度下对应的 Lateral Alert Limit 和 Longitudinal Alert Limit 的值，如表 7 所示。

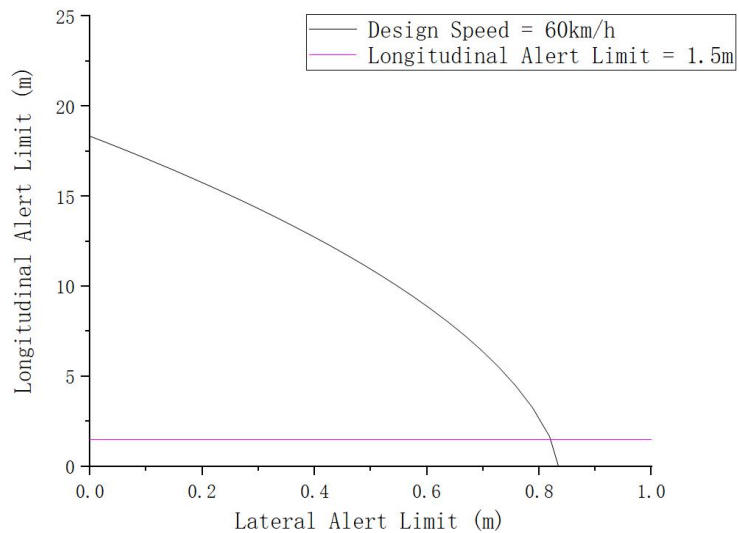

图 5 设计速度为 60km/h 时的 Lateral Alert Limit-Longitudinal Alert Limit 函数曲线

表 7 设计速度与警报限制

| 设计速度 v(km/h) | Lateral Alert Limit/m | Longitudinal Alert Limit/m |
| --- | --- | --- |
| 120 | 0.969315387 | 1.5 |
| 100 | 0.965778949 | 1.5 |
| 80 | 0.960286982 | 1.5 |
| 60 | 0.820757553 | 1.5 |
| 40 | 0.789931446 | 1.5 |
| 30 | 0.607388698 | 1.5 |
| 20 | 0.37227368 | 1.5 |

### 3.3 垂直方向定位警报限制范围计算

在垂直方向上，我们希望能够在复杂的立交桥式的公路结构中（如图 6 的立交桥）确定车辆所在的道路。为实现这一目标，只需 Vertical Alert Limit 的值达到两条道路间垂直高度差最小值的 1/3，即：

$$\text{Vertical Alert Limit} = \frac{\min \Delta \text{Vertical Height}}{3} \quad (7)$$

而基于道路建筑限界（boundary line of road construction）的要求，两条道路间的垂直高度差需大于等于处在下方的道路的净空高度，即处在下方的道路的净空高度可作为道路间垂直高度差的最小值。同时，中国公路净空高度的规定为：高速公路、一级公路、二级公路的净空高度应为 5.00m，三级公路、四级公路的净空高度为 4.50m。[13]考虑到存在上下两条道路类型不一致的情况，选择三级公路、四级公路的净空高度即 4.50m 作为道路间垂直高度差的最小值，此时得到：

$$\text{Vertical Alert Limit} = \frac{4.5\text{m}}{3} = 1.5\text{m}$$

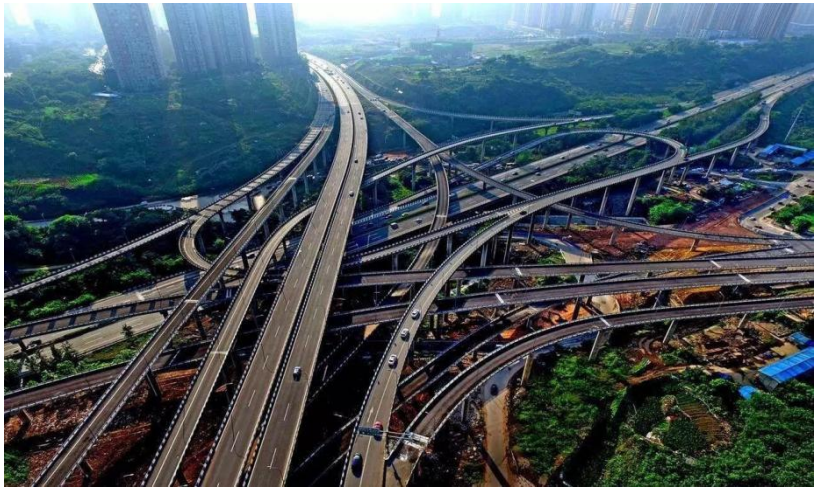

图 6 重庆黄桷湾立交桥，高达 5 层、共 20 条匝道[16]

## 4. 基于中国道路的自动驾驶定位精度需求分析

在得到具体风险分配下对警报限制值要求后，结合位置定位的期望分布与要求，接下来便可以对定位精度需求进行研究，从而整理出不同车辆级别对应的需求设计。

在统计学中，误差常常用正态分布来描述，因此可以认为定位时产生的误差也是近似满足正态分布的。那么，在误差满足正态分布的情况下，之前计算得到的定位(localization)准确度 $(1-10^{-9})$ 即 99.9999999%所对应的置信区间为 $[-6.11\sigma, 6.11\sigma]$。相比之下，在 ISO26262 中使用的 95%准确度对应的置信区间为$[1.96\sigma, 1.96\sigma]$[11]。99.999999%准确度的置信区间是 95%准确度的置信区间的 6.11/1.96=3.12 倍。同时，99.999999%准确度对应的是车辆的警戒边框（Bounding Box）在各个方向上的最大允许误差即 alert limit。因此，根据置信区间的意义，95%准确度对应的各个方向上的误差应该为最大允许误差即 alert limit 的 1/3.12。

接下来需要确定精确控制时定位问题的几何结构，以确定定位限界。这将显示为道路几何标准的函数，如车道宽度和道路曲率以及允许的车辆尺寸。美国关于这一分析将集中在美国的道路和乘用车标准上[11]，本文则以中国的中型轿车尺寸和高速公路车道尺寸来作为研究对象，因为同样的原则也适用于其他地区和车辆类型。在安全关键定位系统中，位置最大可能误差的瞬时估计称为保护等级。本文对车辆周围横向、纵向和垂直保护等级的视为一个"盒子方框"，再定义允许的横向、纵向和垂直保护级别的硬边界，这个边界值称为警报限值，有了这些这些限值，才能确定车辆是否位于车道内。如果设计的防护等级超过警戒极限，将不能保证车辆在车道内。本文推导了基于目标安全级别 TLS 的系统完整性风险分配。然后，按照民用航空中开发的方法，将风险预算分配到整个自主车辆系统中，即如图 1。本文定义

问题的几何结构，根据车辆尺寸和道路几何结构建立定位界限，结合这些，定义了位置误差的期望分布和定位要求。

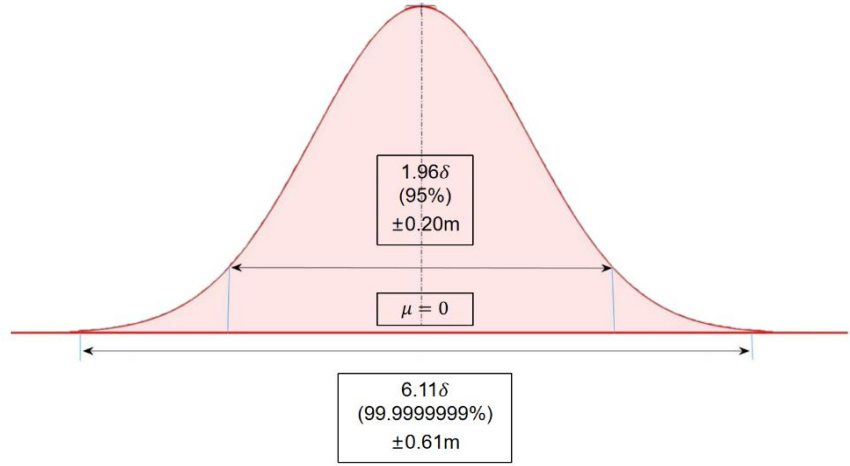

图 7 高斯概率分布

基于第三部分计算的道路车辆尺寸与"定位盒子"边界的关系，本文选取的公路宽度 w 为 3.5 米，曲率 r 为 125 米，车道设计车速是 60km/h，得到关于横向与纵向警报的曲线图，同时限制纵向警报小于等于车长一半或者 1.5 米。本文研究的车辆选择中型轿车，车长 $L_v$ 为 4.7 米，车宽 $W_v$ 为 1.8 米，由图中可以得出横向警报为 0.82 米。最后基于公式

$$\text{Lateral Alert Limit} = (x - W_v)/2, \quad \text{Longitudinal Alert Limit} = (y - L_v)/2 \qquad （8）。[9]$$

计算得出定位盒子横向边界 x 是 3.3 米，纵向定位边界 y 是 9.4 米或 7.7 米。接下来计算定位精度，需要平衡位置与姿态的误差，保证保护级别要小于警报限制值。对于横向、纵向和垂向的方向角，本文假设这 3 个角度的误差相同，那么进一步得到横向、纵向以及垂向的设计公式:

$$\delta\text{lat} + (\delta\text{lon} + \delta\text{vert} + l_v/2)\delta\lambda \leq \text{Lat.AL},$$
$$\delta\text{lon} + (\delta\text{lat} + w_v/2)\delta\varphi + \delta\text{vert}\delta\varphi \leq \text{Lon.AL} \qquad （9）$$
$$\delta\text{vert} + (\delta\text{lat} + w_v/2)\delta\theta + (\delta\text{lon} + l_v/2)\delta\varphi \leq \text{VAL}$$

并在横向上得到一个很好的经验误差公式：$\delta\text{lat} + l_v\delta v \leq \text{Lat.AL}$。[9] （10）

由以上的数据可知，Lat.AL 为 0.82 米。研究的车长为 4.7 米，$\delta\lambda$ 的合理误差需小于 0.1 弧度，否则很快就会超过这个限值，此处取为 0.03 弧度，代入得横向定位误差为 0.677 米，即横向定位精度为 0.216 米。再结合之前有关 95%准确度和 99.999999%准确度之间的关系，车辆在 0.22 米（95%）的定位精度范围内，硬边界误差为 0.63 米（99.9999999%）。

而纵向及垂向定位精度要求则更为宽泛一些，故对结果进行一些小范围近似模糊处理。计算统计了不同尺寸轿车的在允许范围内的定位误差，同样统一选择 1.5 米，则横向警报限值为 0.82 米，并且为了满足保护等级小于警报等级，三个方向角度误差均选为 0.03 弧度以保证定位要求。取角度误差 0.02 弧度（95%），垂直方向 95 百分位净高为 1.5 米。此外每一种车型车长和车宽均选择该类车型范围内最大尺寸。

表 8 车辆级别与定位精度

| 车辆级别 | 车长（总长）/米 | 车宽/米 | 横向定位精度/米 | 纵向定位精度/米 | 垂向定位精度/米 |
|---|---|---|---|---|---|
| *A00* | 3.7 | 1.675 | 0.227 | 0.47 | 0.458 |
| *A0* | 4.4 | 1.75 | 0.22 | 0.47 | 0.458 |
| *A* | 4.7 | 1.825 | 0.217 | 0.47 | 0.458 |
| *B* | 5 | 1.9 | 0.214 | 0.47 | 0.458 |
| *C* | 5.1 | 1.94 | 0.213 | 0.47 | 0.458 |
| *D* | 5.3 | 1.975 | 0.211 | 0.47 | 0.458 |

## 5. 结论

本文研究计算显示，中国的道路安全目标等级 TLS 数量级为 $10^{-11}$，安全等级完全达到了 ASIL-D，其安全性毋庸置疑。此外，基于安全目标等级的要求和虚拟系统风险分配导图，逆向推导计算了各类车辆所必需的定位精度。之后本文以中型轿车为研究对象，计算了它在设计时速 60km/h 的高速公路或立交桥上的定位误差横向、纵向、垂向警报限制值分别为 0.82m（95%）、1.5m（95%）、1.5m（95%），横向，纵向，垂向定位精度需求是 0.203m（95%）、0.488m（95%）、0.45m（95%）。然后统计计算了其他道路情况下的警报限制值和其他车型的定位精度设计需求。在各种道路条件下，这个定位精度（95%）是一个警报限制，也是保护等级下限。本文计算了车辆定位精度需求，目的是在应用定位设备时有一个参考数据，更好的去进行自动驾驶设计。但是，由于中国道路情况太过于复杂，本文对一些乡村公路、县道以及一些复杂曲率的道路等并未深入计算，这有待车辆设计者具体实验后针对设计。同时，更多道路类型运行数据也有待积累，以便为自动驾驶车辆或道路设计提供支持。

# 参考文献